\documentclass[12pt]{article}
\usepackage{scicite}
\usepackage{times}
\usepackage{graphicx}
\usepackage{amssymb}

\newenvironment{sciabstract}{%
\begin{quote} \bf}
{\end{quote}}

\newcounter{lastnote}
\newenvironment{scilastnote}{%
\setcounter{lastnote}{\value{enumiv}}%
\addtocounter{lastnote}{+1}%
\begin{list}%
{\arabic{lastnote}.} {\setlength{\leftmargin}{.22in}}
{\setlength{\labelsep}{.5em}}} {\end{list}}

\title{Is the Fragility of a Liquid Embedded \\ in the Properties of its Glass?}

\author
{Tullio Scopigno,$^{1\ast}$ Giancarlo Ruocco,$^{1}$ Francesco Sette$^{2}$, Giulio Monaco$^{2}$\\
\\
\normalsize{$^{1}$INFM and Dipartimento di Fisica, Universit\'a di Roma La Sapienza, 00185 Roma, Italy,}\\
\normalsize{$^{2}$European Synchrotron Radiation Facility, BP 220, 38043 Grenoble, France}\\
\normalsize{$^\ast$To whom correspondence should be addressed;
E-mail:  tullio.scopigno@phys.uniroma1.it} }

\date{}

\begin{document}

\baselineskip24pt

\maketitle


\begin{sciabstract}
When a liquid is cooled below its melting temperature it usually
crystallizes. However, if the quenching rate is fast enough, it is
possible that the system remains in a disordered state,
progressively losing its fluidity upon further cooling. When the
time needed for the rearrangement of the local atomic structure
reaches approximately $100$ seconds, the system becomes "solid"
for any practical purpose, and this defines the glass transition
temperature $T_g$. Approaching this transition from the liquid
side, different systems show qualitatively different temperature
dependencies of the viscosity, and, accordingly, they have been
classified introducing the concept of "fragility". We report
experimental observations that relate the microscopic properties
of the {\it glassy phase} to the fragility. We find that the
vibrational properties of the glass {\it well below} $T_g$ are
correlated with the fragility value. Consequently, we extend the
fragility concept to the glassy state and indicate how to
determine the fragility uniquely from glass properties well below
$T_g$.
\end{sciabstract}


When a liquid is cooled, the loss of kinetic energy leads to an
ordering of the molecules which then crystallize at the melting
temperature $T_m$. However, if cooled fast enough through $T_m$,
some materials (glass forming materials) are capable to sustain a
metastable liquid state and, upon further cooling, to freeze into
a disordered glassy state \cite{creta,pisa1999,pisa2003,db}. The
law that describes the change of the viscosity with the
temperature approaching the glass transition temperature, $T_g$,
is highly material specific, and has led to the classification of
the glass formers materials according to the concept of
"fragility" \cite{ang,ang2}. The kinetic fragility, $m$, is
directly related to the slowing down of the dynamics: it is
defined in terms of the shear viscosity $\eta$ as:

$$m = \lim_{T \rightarrow T_g}  \frac{d \log(\eta)}{d(T_g/T)}$$

Therefore, $m$ is an index of how fast the viscosity increases
approaching the structural arrest at the glass transition
temperature $T_g$, the temperature where the structural
relaxation time $\tau_\alpha \sim 100$ s. At this latter
temperature, through the Maxwell relation $\eta=G_\infty \tau$,
corresponds a viscosity $\eta \sim 10^{13}$ poise (more likely
$\eta \sim 10^{11}$ poise for molecular glasses), while $10^{-4}$
poise is the "infinite" temperature limit in basically any
materials. Consequently, fragility values typically range between
$m=17$ for "strong" systems -those that show an Arrhenius
behaviour- and $m \sim 150$ for "fragile" systems, where the high
cooperativity of the diffusive dynamics induces a high (and
$T$-dependent) apparent activation energy. One interest in this
classification lies in the attempt to relate the temperature
behavior of a macroscopic transport property close to $T_g$ to
the microscopic interactions driving the dynamics of the system.
It has been found, for example, that the value of the fragility is
empirically related to the kind of interaction potential among
the particles constituting the system. Prototypical examples of
fragile liquids are those composed by units interacting via
isotropic bonds, such as Van der Waals-like molecular liquids.
The strong glass-forming liquids, on the other hand, are those
characterized by strong covalent directional bonds that form space
filling networks. O-terphenyl ($m$=80) and $SiO_2$ ($m$=20) are
characteristic examples of a fragile and a strong liquid,
respectively. Hydrogen bonded systems, as glycerol ($m$=50), are
often called "intermediate" between strong and fragile liquids.
The kinetic fragility has been found to be correlated to other
properties of the glass-forming liquids, such as: i) the slope at
$T_g$ of the temperature dependence of the configurational
entropy (often referred to as thermodynamic fragility)
\cite{mar}, or -for classes of systems sharing similar glass
transition temperatures- the specific heat jump at $T_g$
\cite{huang_ter,ngai_ter}; ii) the ratio between the maximum and
the minimum of the boson peak, i.e. of the bump observed at the
THz in the Raman and neutron scattering spectra of glass-forming
materials \cite{sok_fbp}, but this finding is controversial
\cite{yann_fbp}; iii) the degree of stretching in the
non-exponential decay of the correlation functions in the liquid
close to $T_g$ \cite{boh_data}, iv) the statistics of the minima
in a potential energy landscape-based description
\cite{land1,land2} of the diffusion process in supercooled
liquids \cite{waly_frag,spe,sas}, and, more recently, v) the
temperature behaviour of the shear elastic modulus in the
supercooled liquid \cite{dyr}. In all these studies the fragility
has been always related to (or defined through) macroscopic
properties characterizing the liquid side of the
glass-transition. While there are attempts to relate the
fragility to the anharmonicity of the "hot" glass
\cite{ang_school}, no connection has been found up to now between
the value of $m$ and the physical properties of the low
temperature glassy phase.

We show that, starting from a determination of the non ergodicity
factor in the low temperature glass, it is possible to identify a
parameter that controls how fast the non ergodicity factor
decreases on increasing the temperature, and that turns out to be
proportional to the fragility $m$. Through this, we establish a
way to determine the fragility of a system in the glassy phase
well below $T_g$, independent of the way the viscosity changes
with decreasing temperature from the liquid side. By exploiting
the harmonic approximation of the low temperature dynamics, it is
found that this parameter only depends on the characteristics of
the static disorder, which, in turn, controls the vibrational
eigenmodes of the glass. This result demonstrates the existence of
a deep link between the diffusive inter-basins dynamics and the
vibrational intra-basin dynamics.

Recent extensive inelastic x-ray scattering (IXS) measurements of
the dynamic structure factor have allowed to constitute a
sizeable library of high frequency (THz) dynamical properties of
glasses. Of interest here, the IXS measurements allow for the
determination of the non-ergodicity factor, $f(Q,T)$ with a
reliability that was not achievable via light or neutron
scattering \cite{set_sci}. The non ergodicity factor is the long
time limit of the density correlator, $\Phi_Q(t)$, i.~e. the
density-density correlation function, $F(Q,t)$, normalized to the
static structure factor, $S(Q)$: $\Phi_Q(t)=F(Q,t)/S(Q)$. The
quantity 1-$f(Q,T)$ represents the amount of decorrelation
introduced by the vibrational dynamics, and it depends on both the
($T$-dependent) amplitude of the vibrations and the degree of
disorder of the glassy structure. In a low temperature glass,
$F(Q,t)$, apart from the Debye-Waller factor $\exp{(-W(Q))}$, can
be expressed as the sum of a constant term $S_{IS}(Q)$, which
represent the static structure factor of the atomic equilibrium
positions (Inherent Structure), plus a time-dependent one,
$F_{inel}(Q,t)$, which is the contribution of the atomic
vibration around such equilibrium positions, a quantity that
vanishes in the long time limit:

\begin{equation}
F(Q,T)= e^{-W(Q)}\left [ S_{IS}(Q)+F_{inel}(Q,t)\right]
\label{Fqt}
\end{equation}

\noindent Therefore:

\begin{equation}
f(Q,T)=\lim_{t \rightarrow \infty} \Phi_Q (t) = \lim_{t
\rightarrow \infty}
\frac{S_{IS}(Q)+F_{inel}(Q,t)}{S_{IS}(Q)+S_{inel}(Q)} =
\frac{1}{1+S_{inel}(Q)/S_{IS}(Q)} \label{Fqt1}
\end{equation}

\noindent where we have defined $S_{inel}(Q) \doteq
F_{inel}(Q,t=0)$. $S(Q,\omega)$ is the Fourier transform of
$F(Q,t)$ and is the quantity directly accessible in scattering
experiments. From Eq.(\ref{Fqt}) it can be expressed as

\begin{equation}
S(Q,\omega)=e^{-W(Q)}\left [
S_{IS}(Q)\delta(\omega)+S_{inel}(Q,\omega)\right] \label{sqw}
\end{equation}

From an experimental point of view and according to Eqs
(\ref{Fqt1}) and (\ref{sqw}), the non ergodicity factor is
derived from the ratio of the elastic to the inelastic scattered
intensity, obtained from inelastic scattering measurements of the
dynamic structure factor $S(Q,\omega)$ \cite{nota}. A sense of the
$T$ dependence of $f(Q,T)$ can be obtained from Fig.~1. Here, as
an example, we report the IXS spectra at fixed exchanged
wavevector ($Q=2$ nm$^{-1}$) and at different temperatures in
glycerol. The inelastic (dashed lines) and elastic (dotted lines)
contributions to the scattering intensity are shown and one can
appreciate in the raw data the change of relative intensity as a
function of $T$. As far as the $Q$ dependence is concerned,
$f(Q,T)$ follows in phase the oscillations of the static
structure factor and is almost $Q$-independent in the
$Q\rightarrow 0$ region where $S(Q)$ is almost constant
\cite{tolle_otp}, see the inset of Fig.~2. We focus on this
small-$Q$ region. From the integrated intensities of the elastic
and inelastic contributions, obtained by a fitting procedure (see
Eq.~(S1) and (S2) of the supporting on line materials), the $T$
dependence of $f(Q,T)$ is obtained. The values of $f(Q,T)^{-1}$
(which is expected to be linear in $T$, see below) are reported in
Fig.~2 (triangles). Also reported in the same figure is the $T$
dependence of $f(Q,T)$ for two other archetypical glasses: silica
and o-terphenyl (oTP).

To better understand the temperature dependence of $f(Q,T)$ in the
$T\rightarrow 0$ limit, we invoke the harmonic approximation for
the vibrational dynamics. This allows one to express $f(Q,T)$ in
terms of the vibrational properties of the systems, i.~e. the
eigenvalues ($\omega_p$) and eigenvectors ($\bar e_p$) of the the
potential energy Hessian evaluated at the inherent structure.
Using the harmonic approximation for $S_{inel}(Q,\omega)$, it is
straightforward \cite{gcr_prlsim,scop_presim} to show that
Eq.(\ref{Fqt1}) reduces to:

\begin{equation}
f(Q,T)=\left [1+\frac{K_B T Q^2}{M S_{IS}(Q)}\frac{1}{N}\sum_p
{\frac{\left | \sum_i \left [\hat{Q} \cdot e_p (i) \right ]e^{i Q
r_i} \right |^2}{\omega^2_p}} \right ]^{-1}. \label{har}
\end{equation}

Here $M$ is the molecular mass, $K_B$ the Boltzmann constant, and
$i$ is summed over the $N$ particles and $p$ over the $3N$ normal
modes. In order to pinpoint the $T$ dependence of the non
ergodicity factor in the low $Q$ limit it is convenient to rewrite
Eq.~(\ref{har}) as

\begin{equation}
f(Q\rightarrow 0,T)=\frac{1}{1+\alpha \frac{T}{T_g}}. \label{har2}
\end{equation}

\noindent We thus define the dimensionless quantity $\alpha$,
which encompasses all the microscopic details of the system, as
the eigenvalues and eigenvectors of the normal modes. These are
quantities that, in turn, depend on the interaction potential and
on the disordered structure. This equation provides a formal way
to extract the system-dependent parameter $\alpha$ from the $T$
dependence of $f(Q,T)$, derived from the IXS data. This has
motivated us to revisit the large amount of IXS data available
for glasses at low $T$ where the harmonic approximation, and
therefore Eq.(~\ref{har2}), is expected to be valid. As it can be
seen in Fig. 2, the observed $T$ dependence of $f(Q,T)$ is fully
consistent with the functional form predicted by Eq.(~\ref{har2}),
and this allows us to determine $\alpha$ by a least square
minimization procedure. The derived values for $\alpha$ (e.~g.
$\alpha$=0.19 for silica, $\alpha$=0.32 for glycerol and
$\alpha$=0.58 for oTP) clearly indicate a trend: the more fragile
a liquid, the greater the slope of $f(Q,T)$ at $T$=0, i.~e. the
faster the decorrelation of the density fluctuations on
increasing $T$. The fitting procedure has been applied to the
whole set of available glasses, and the obtained values of
$\alpha$ are reported in Tab.I and Fig.~3 as a function of the
independently known fragility parameter $m$. From this figure
clearly emerges the existence of a strong correlation between $m$
and $\alpha$: the higher the fragility, the higher the value of
$\alpha$, i.~e. the faster is the $T$ dependence of the $f(Q,T)$
parameter. The existence of a strong correlation between $\alpha$
and $m$ is further emphasized by the empirical observation that
the two quantities are not only correlated but (within the
statistical accuracy) actually proportional to each other,
according to the relation $m=(135\pm 10)\alpha + (4\pm 5)$. On
passing, we note that the two points that lie definitely below
the dotted line (Selenium and Salol) are the ones for which
fragility determined at $T_g$ doesn't agree well with the
fragilities determined at higher temperatures
\cite{mar,garr_frag}.

The observed correlation is conceptually surprising. It indicates
the existence of a link between the curvatures of the potential
energy function at its minima (more specifically those visited in
the glassy phase) and the other properties of the potential
energy function (energy distribution of the minima,
minimum-to-minimum barrier height, distribution of the saddle
order and energies,...) controlling the diffusive processes in
supercooled liquids.

We further examine how $\alpha$ emerges from the collective
density-density correlation function plateau. From Eqs.
(\ref{har}) and (\ref{har2}), the microscopic expression for
$\alpha$ is found to be:

\begin{equation}
\alpha=\frac{K_B T_g Q^2}{M S_{IS}(Q)}\frac{1}{N}\sum_p
{\frac{\left | \sum_i \left [\hat{Q} \cdot e_p (i) \right ]e^{i Q
r_i} \right |^2}{\omega^2_p}} \label{alpha}
\end{equation}

One may in principle derive a similar parameter $\alpha_s$ from
the temperature dependence of the self correlator plateau. In this
case $\alpha_s$ is related to the familiar mean square
displacement.

\noindent The harmonic approximation for the Debye Waller factor
$f_s(Q,T)=\exp{(-W(Q,T))}$ would lead to an equation formally
identical to Eq.(\ref{har2}), but with $\alpha$ replaced by
$\alpha_s$:

\begin{equation}
\alpha _s=\frac{K_B T_g Q^2}{M}\frac{1}{N}\sum_p
{\frac{1}{\omega^2_p}}
\end{equation}

\noindent Therefore $\alpha$ and $\alpha_s$ differently weight
the low frequency modes. Specifically, in the small $Q$ limit,
$\alpha$ is more sensitive to the low energy modes than
$\alpha_s$ that, independently on $Q$, reflects the whole density
of states. It would be interesting to check whether or not these
two quantities are correlated with each other.

In conclusion, we report evidence for the correlation between the
fragility of a glass-forming liquid and the temperature
dependence of its non-ergodicity factor as determined by the
vibrational dynamics at very low temperatures. The fragility is
an index of how the viscosity increases upon supercooling. The
non ergodicity factor, in the low temperature limit, is related
to the vibrational properties of the harmonic glassy dynamics
(see Eq.(\ref{alpha})), i.e. to the curvature of the energy
minima. Therefore, from our finding it emerges that the
properties of the potential energy landscape around the deep
minima are related to those properties that control the diffusion
of the system through different basins. This unexpected relation
represents a further aspect that requires to be clarified in the
physics of the glass-transition.

\newpage
\begin{table}[h]
\centering
\begin{tabular}{|c|c|c|c|} \hline\hline Sample&$T_g [K]$&$m$&$\alpha$\\ \hline
$\mathrm{BeF_2}^a$&598&20&0.16\\
silica$^b$&1450&28&0.19\\
glycerol$^c$&190&53&0.32\\
$\mathrm{PB_{1.4}}^d$&180&60&0.40\\
nBB$^e$&125&53&0.46\\
salol$^f$&218&73&0.64\\
mtol$^g$&187&77&0.57\\
oTP$^h$&241&81&0.58\\
mTCP$^i$&205&87&0.59\\
Se$^j$&308&87&0.7\\
\hline\hline
\end{tabular}
\caption{Temperature steepness of the viscosity at $T_g$
(fragility) and of the non ergodicity factor at $T\rightarrow 0$
($\alpha$) for several materials.
\newline $^a m$ from Ref.\cite{hem_bef2},
$\alpha$ from Ref.\cite{sco_bef2}.
\newline $^b$Silica infrasil grade sample. $m$ from
Ref.\cite{boh_data}, $\alpha$ from Ref.\cite{mas_17nomi} and
references therein.
\newline $^c m$ from Ref.\cite{boh_data}, $\alpha$ from
Ref.\cite{ruo_gly}.
\newline $^d$Polybutadiene: $\alpha$ is determined from an experiment performed
on a (1.2PBD)$_0$(1.4PBD)$_{100}$ reported in Ref.\cite{fio_pb}.
The fragility is not available for such concentration, we estimated
$m$ extrapolating data from
Refs.\cite{zor_pb,boh_data}.
\newline $^e$ Normal-butyl-benzene: $m$ from Ref.\cite{han_nbb}, $\alpha$ from the experiment in
Ref.\cite{mas_glasses}.
\newline $^f m$ from Ref.\cite{boh_data},
$\alpha$ from unpublished data.
\newline $^g$Meta-toluidine: $m$ from Ref.\cite{erw_mtol}, $\alpha$ from unpublished
data.
\newline $^h$ Orto-terphenyl: $m$ from Ref.\cite{boh_data}, $\alpha$ from
Ref.\cite{mon_otpfq}.
\newline $^i$ Meta-tricresyl-phosphate: $m$ from Ref.\cite{boh_data}, $\alpha$ from unpublished
data.
\newline $^j m$ from Ref.\cite{boh_data}, $\alpha$ from unpublished
data.
} \label{tab1}
\end{table}

\newpage

\begin{thebibliography}{10}
\bibitem{nota}
Some fraction of the elastic line may be in principle related to
the
  non-propagating temperature fluctuations. The amount of this contribution is
  actually ruled out by the quantity $\frac{c_p}{c_v}-1=\gamma-1$ and therefore
  it vanishes in the $T \rightarrow 0$ limit when $\gamma \rightarrow 1$ holds
  according to the harmonic approximation.
\end{thebibliography}

\begin{scilastnote}
\item The authors gratefully acknowledge C.A. Angell for valuable
hints and suggestions. J. Dyre, S. Sastry, F. Sciortino and S.
Yannopoulos are also acknowledged for fruitful discussions.
\end{scilastnote}

\newpage
\begin{figure} [f]
\centering
\includegraphics[width=.8\textwidth]{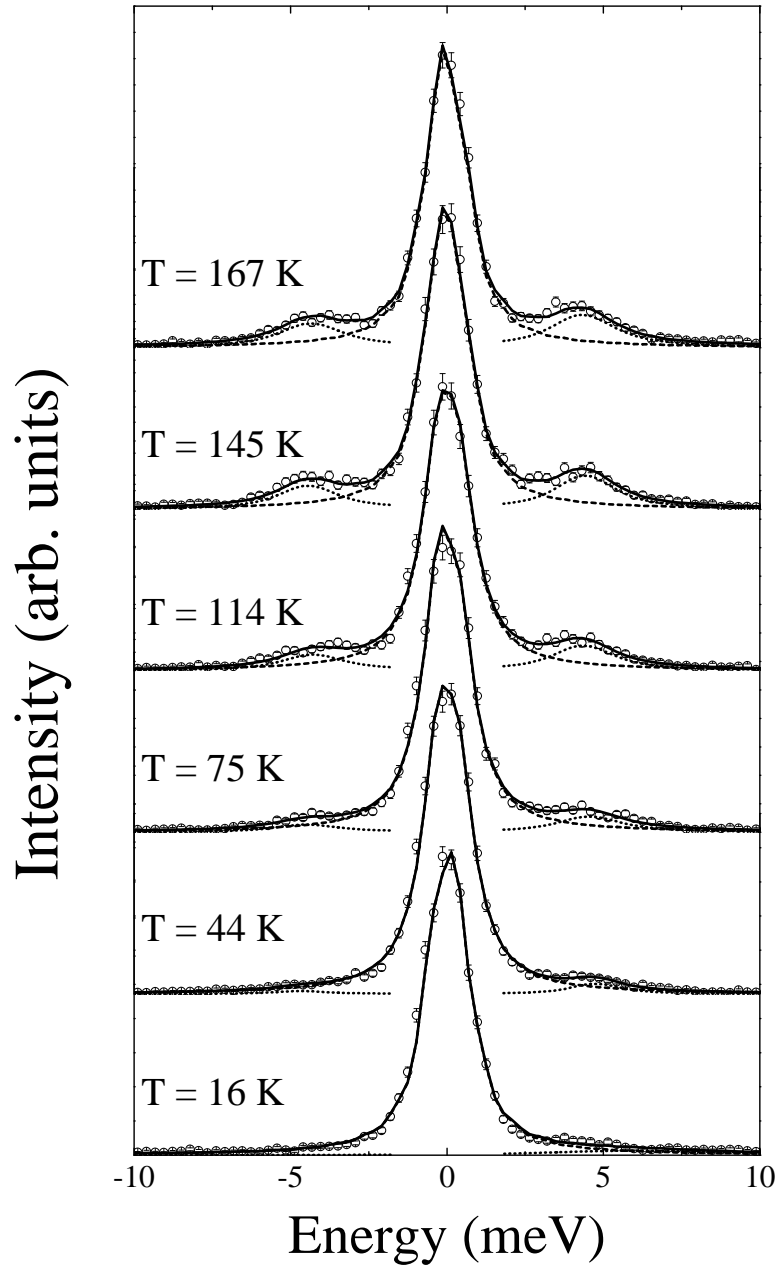}
\vspace{0 cm} \caption{Selected example of the inalastic x-ray
scattering energy spectra of glycerol (open circles with error
bars) taken at $Q=2$ nm$^{-1}$ at the indicated temperatures. The
solid line is the line of best fit according to Eqs (\ref{sqw})
and (S2), while the dashed and dotted lines are the elastic and
inelastic contributions, respectively (see the supporting on line
material for further details). The values of $f(Q,T)$ are obtained
by the ratio of the integrated intensities of the elastic and
inelastic contribution, from Eq.(\ref{Fqt1}).} \label{final}
\end{figure}

\newpage
\begin{figure} [f]
\centering
\includegraphics[width=1.\textwidth]{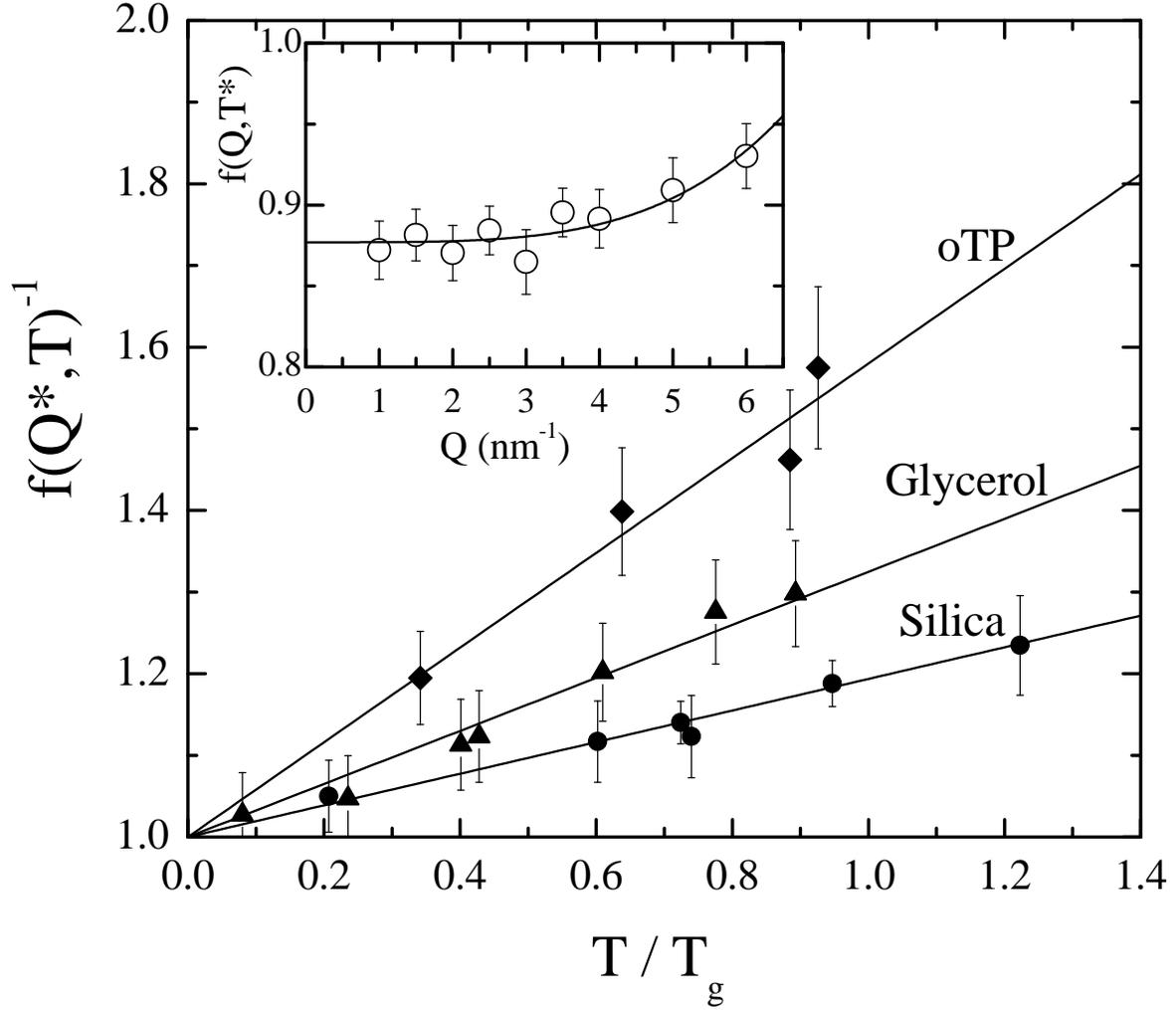}
\vspace{-9 cm} \caption{Values of $f(Q^*,T)^{-1}$
(Eq.(\ref{Fqt1})) for $Q^*\simeq 2$ nm$^{-1}$ reported in a
$T/T_g$ scale for three representative materials (full symbols
with error bars): silica ($T_g$=1450 K), glycerol ($T_g$=190 K)
and oTP ($T_g$=241 K). The full line is the best fit of the
experimental data to Eqs (\ref{har2}). These fits have been used
to derive the values of $\alpha$ reported in Fig.~3. In the inset
we show the $Q-$dependence of $f(Q,T^*)$ for Silica at $T^*=1050$
K} \label{final}
\end{figure}

\newpage
\begin{figure} [f]
\centering
\includegraphics[width=1.\textwidth]{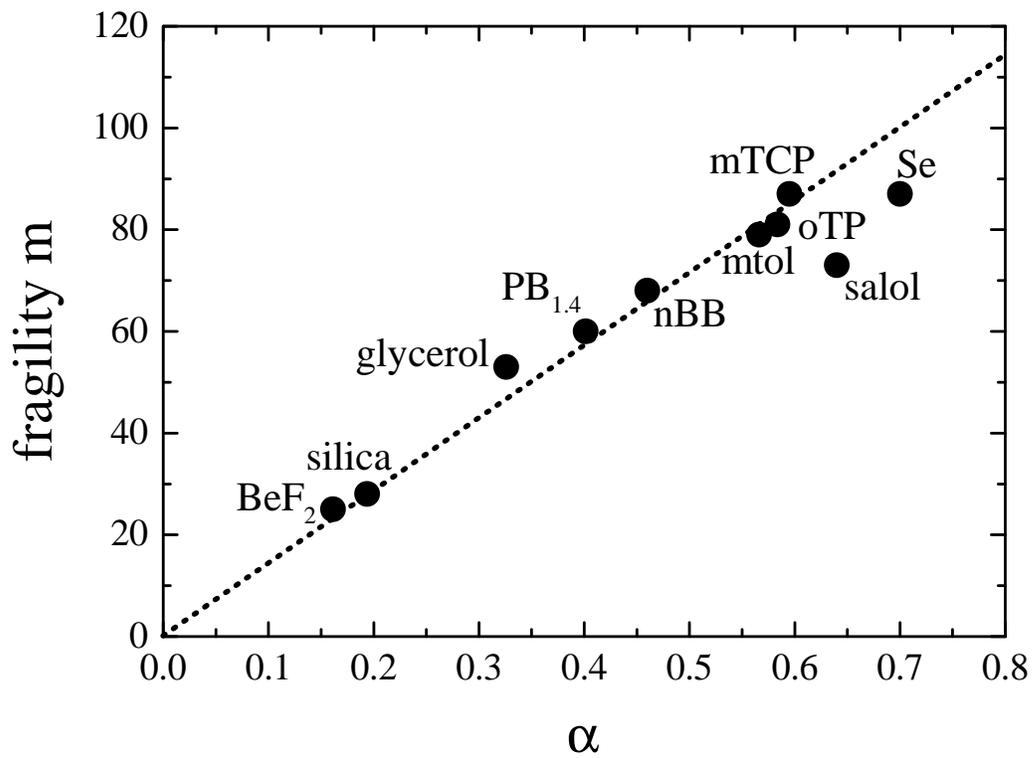}
\vspace{-11.5 cm} \caption{Correlation plot of the kinetic
fragility and the $\alpha$ parameter of the non-ergodicity factor
(see Eq.~(\ref{har2})). The dotted line is obtained by a fit of
the data to a linear equation. It corresponds to $m=135\alpha$ and
the regression coefficient is $r$=0.994.} \label{final}
\end{figure}

\end{document}